\documentclass[prb,showpacs,superscriptaddress,twocolumn,aps]{revtex4-1}
\usepackage{graphicx}
\usepackage{amsmath}
\usepackage{amssymb}
\usepackage{hyperref}\hypersetup{colorlinks=true,linkcolor=blue,citecolor=blue,pdfstartview=FitH}
\usepackage{bm}\renewcommand{\mathbf}{\bm}
\usepackage{xcolor}
\definecolor{blue2}{cmyk}{0.77,0.51,0.01,0}

\DeclareMathOperator{\Tr}{Tr}

\DeclareMathOperator{\Real}{Re}
\DeclareMathOperator{\Imag}{Im}\renewcommand{\Im}{\Imag}

\usepackage[caption=false]{subfig}

\begin{document}
\title{Effective temperature from fluctuation-dissipation theorem in systems with bipartite eigenmode entanglement}
\author{T.S.~Bortolin}
\author{A.~Iucci}
\affiliation{Instituto de F\'{\i}sica La Plata (IFLP) - CONICET}
\affiliation{Departamento de F\'{\i}sica - Universidad Nacional de La Plata, CC 67, 1900 La Plata, Argentina}

\begin{abstract}
In thermal equilibrium, the fluctuation-dissipation theorem relates the linear response and correlation functions in a model and observable independent fashion. Out of equilibrium, these relations still hold if the equilibrium temperature is replaced by an observable and frequency-dependent parameter (effective temperature). When the system achieves a long time thermal state all of these effective temperatures should be equal and constant. Following this approach we examine the long times regime after a quantum quench in a system with bipartite entanglement in which the asymptotic values of the observable are compatible with the ones obtained in a Gibbs ensemble. We observe that when the initial entanglement is large, and for a large range of (intermediate) frequencies, the effective temperatures corresponding to the analyzed local and non-local operators approach an approximate constant value equal to the temperature that governs the decay of correlations. Still, the residual frequency dependence in the effective temperature, and the differences observed among observables discards strict thermalization.
\end{abstract}

\maketitle

\section{Introduction}
\label{sec:intro}

A series of experiments with ultra cold atoms carried out in the last decade
\cite{greiner02_fast_tunnability,kinoshita06_non_thermalization,hofferberth07_nonequilibrium_dynamics_1D_bosons,trotzky11_relaxation_BH_10101010,chenau12_light_cone_experimental_bosons,gring12_pre-thermalization_isolated_bose_gas,langen13_local_equilibrium} exhibited absence of dissipation in the many-particle system and therefore essentially unitary time evolution on long time scales. This motivated a great deal of activity involving the study of the dynamics of interacting quantum systems that are driven out of equilibrium by preparing them in an initial state that is not in the eigenbasis of the Hamiltonian. Several interesting problems arise in these systems such as the thermalization mechanisms in integrable and non-integrable models~(see Refs.~\onlinecite{cazalilla10_quenches,dziarmaga10_quenches,polkovnikov11_nonequilibrium_dynamics} and references therein) and more generally the emergence of thermodynamics in isolated systems.

Much of the theoretical effort has been devised to investigate exactly solvable models and integrable systems, which are special since the large number of integrals of motion that constrain the nonequilibrium dynamics are believed to preclude the relaxation to thermal equilibrium. Instead, in many cases the long-times steady state is captured by a statistical description based on a generalized Gibbs ensemble (GGE)~\cite{rigol07_generalized_gibbs_hcbosons} which results from the maximization of the entropy subjected to the constraints imposed by the conserved quantities. In such a description a different temperature is associated with each conserved quantity.

Interestingly, it was shown in Refs.~\onlinecite{rigol06_hcbosons_dynamics,rigol11_quench_initial_state_dependence}
that certain kinds of initial states can lead to asymptotic values of the observables whose GGE description is essentially indistinguishable from the one computed with a standard thermal Gibbs ensemble. This effect turns out to be generic for integrable models that can be mapped onto quadratic, bosonic or fermionic models and initial states for which two sets of modes are strongly entangled~\cite{chung12_thermalization_entanglement}. However, the GGE cannot reproduce the behavior of all observables~\cite{iucci09_quench_LL}, and in particular it fails to capture energy fluctuations. Therefore, the effective temperature that emerges from the standard Gibbs distribution description characterizes the asymptotic thermal correlations and constitute a measure of the entanglement between the eigenmodes in the initial state, but does not have the usual thermodynamic meaning.

One important relation in equilibrium statistical mechanics both quantum and classical is the Fluctuation-Dissipation theorem (FDT), that relates linear response and correlation functions in a model and observable-independent fashion. Even though the FDT is strictly valid for systems in thermodynamic equilibrium, in many out-of-equilibrium situations, the FDT turns out to be more relevant for the analysis of thermalization issues than the functional decay of observables~\cite{cugliandolo11_effective_temperature_slow_dynamics}. It was shown to hold out of equilibrium after relaxation, in both nonintegrable~\cite{essler12_dynamical_gge,khatami13_FDT_noneq_nonintegrable} and integrable~\cite{essler12_dynamical_gge} systems. However, in the latter case only a basic form of it holds, implying that the way in which deviations from equilibrium states originated in external perturbations and random fluctuations dissipate in time are related, but a detailed balancing relation between the probabilities of energy absorption and release involving only the temperature of the system breaks down. Still, it is possible to define an effective temperature from the FDT~\cite{cugliandolo11_effective_temperature} in the context of quantum quenches as was done for example in integrable models such as the Luttinger model~\cite{mitra11_quench_mode_coupling,mitra12_quench_thermalization_dissipation} and the transverse field Ising chain~\cite{foini11_quench_FDT,foini12_quench_FDT_long}. The effective temperatures defined in this way depend on the momentum and frequency being considered and more important, change according to the observable under study.

In this work we analyze how these ideas apply in the context of a quantum quench for which two sets of modes are strongly entangled in the initial state and that as a consequence exhibits signs of thermalization in the decay of their correlations. We compute dynamic correlation functions of local and non-local operators in a model that is describable in terms of free fermions, from which we extract effective temperatures by forcing the FDT. We show that all the  effective temperatures obtained for local operators have a well defined limit (at least in a certain range of frequencies) when the initial entanglement is strong, that is given by the effective temperature of the system after relaxation. On the other hand, effective temperatures extracted from correlatons of non-local operators exhibit a similar behavior, but its frequency dependence at large values of the initial entanglement show small deviations from that limit.

The rest of this article is organized as follows: In section \ref{sec:model} we present the model (a 1D hard-core boson in presence of a superlattice potential)  and the known results in the generalized Gibbs ensemble. In section \ref{sec:corr}  we study the dynamic two-time correlation functions of Fermi, density and non-local operators. In section \ref{sec:fdr} we introduce the concept of fluctuation-dissipation relations (FDRs) and compute effective temperatures for the operators analyzed in the previous section. In section \ref{sec:conclusions} we present our conclusions and discuss some implications of our work.

\section{The model}
\label{sec:model}

Let us consider a model that describes a system of hard-core bosons in one dimension that initially move in the presence of a superlattice potential. After performing a Jordan-Wigner transformation, this model maps onto the following Hamiltonian
\begin{equation}
H_0=-\sum_j^Lf^\dagger_j f_{j+1}+\mathrm{h.c.}+\Delta\sum_j^L(-1)^j f^\dagger_j f_j, \label{eq:H0}
\end{equation}
written in terms of noninteracting spinless fermions creation $f_j$ and destruction $f_j^\dagger$ operators at site $j$ ($j=1,\ldots,L$, for a lattice of $L$ sites). Periodic boundary conditions (b.c.) in the bosonic model translate into either periodic or antiperiodic b.c. in the corresponding fermionic model depending on whether the number of bosons (fermions) in the system $N$ is odd or even, while open b.c map into open boundary conditions. The system is driven out of equilibrium by preparing it in an initial state  in contact with a thermal reservoir at a temperature $T$ , i.e., it is described by a density matrix $\rho_0 = Z^{-1}e^{H_0/T}$ (such that $\Tr \rho_0 = 1$). For $t > 0$, the superlattice potential is switched off and the system evolves unitarily with a Hamiltonian  $H$ obtained from $H_0$ by setting $\Delta=0$.

Let us first recall the results of Ref.~\onlinecite{chung12_thermalization_entanglement} and show that correlation functions acquire a thermal form for long times.  After Fourier transforming, $H_0$ and $H$ become
\begin{equation}
H_0=H+\Delta\sum_k \left(f^\dagger_{k+\pi}f_k+f^\dagger_k f_{k+\pi}\right)
\end{equation}
and
\begin{equation}
H=\sum_k \omega_k \left(f^\dagger_k f_k-f^\dagger_{k+\pi}f_{k+\pi}\right),
\end{equation}
where $\omega_k=-2\cos k$ and $-\pi/2\leqslant k\leqslant\pi/2$. The existence of the coupling $\Delta$ in $H_0$ implies that in the initial state there are correlations (i.e. bi-partite entanglement) between the eigenmodes at $k$ and $k+\pi$, i.e. $\langle f^\dagger_{k+\pi}f_k\rangle\neq 0$. A Bogoliubov rotation finally renders $H_0$ diagonal with dispersion $E_k=\sqrt{\omega_k^2+\Delta^2}$.

Dephasing makes static correlations at long times to be described by a GGE density matrix that is obtained using the maximum entropy principle taking into account that the system dynamics is constrained by the existence of the set of integrals of motion given by $I_k = f^\dagger_kf_k$ (and $I_{k+\pi} = f^\dagger_{k+\pi}f_{k+\pi}$ (with $k$ restricted to the first Brioulin zone). The GGE density matrix thus obtained reads:
\begin{equation}
\rho_\mathrm{GGE}=\frac{1}{Z_\mathrm{GGE}}\exp\left\{-\sum_k\lambda_k \left( f^\dagger_k f_k-f^\dagger_{k+\pi}f_{k+\pi}\right)\right\},\label{eq:rho_GGE}
\end{equation}
where, at $T=0$ for simplicity,
\begin{equation}
\lambda_k=\log\frac{E_k+\omega_k}{E_k-\omega_k}.
\end{equation}
For $\Delta\gg\omega_k$, $E_k$ can be approximated by $\Delta$ and therefore $\lambda_k=2\omega_k/\Delta$. Thus, the GGE density matrix, equation (\ref{eq:rho_GGE}), reduces to a standard Gibbs ensemble with temperature $T^\mathrm{G}_\mathrm{eff}=\Delta/2$ and the system exhibits thermal correlations.

\section{Dynamic correlations}
\label{sec:corr}

In this section we present our results for the dynamic correlations of several quantities relevant for our model. We  study (anti)symmetrized two-time correlations of two operators $A$ and $B$ in the Heisenberg representation, $A_H(t) = e^{iHt} A e^{-iHt}$,
\begin{equation}
C^{AB}_\pm(t,t_0 )= \langle [A (t+t_0 ),B (t_0 )]_\pm \rangle, \label{eq:corrs}
\end{equation}
where $[X,Y ]_\pm = (XY \pm Y X)/2$ and $\langle \cdots \rangle$ represents the trace over the initial state $\rho_0$. Without loss of generality we consider operators with zero mean value, i.e. $O(t) = O(t)-\langle O(t) \rangle$. We focus on the (anti)symmetric correlator $C_+(C_-)$ and the retarded (or linear response) function, which can be constructed by using $C_\pm$
\begin{equation}
R^{AB}_\pm(t,t_0) = 2i \theta (t) C_\pm^{AB}(t,t_0). \label{eq:resp}
\end{equation}
$R_{\pm}^{AB}$ vanishes for $t<0$ respecting causality. In thermal equilibrium it is related to the correlation function $C_{\pm}^{AB}$ by means of the fluctuation-dissipation theorem (FDT) explained in section \ref{sec:fdr}. While the usual (bosonic) FDT involves $R_-$ and $C_+$, a fermionic version can be constructed by using $R_+$ and $C_-$.   We examine these functions in time domain in section \ref{sec:corrtime}, and in the frequency domain in section \ref{sec:corrfrec}. The latter is in turn used to compute the effective temperature for each pair of operators.

\subsection{Time dependence}
\label{sec:corrtime}
Before starting with the specific two-time correlators calculation, we remark some aspects of the procedure followed and state general results.  We are concerned with the computation of the two-time correlation functions
\begin{equation}
C^A_{\pm,(n,m)}(t,t_0 )= \langle [A_n (t+t_0 ),A_m (t_0 )]_\pm \rangle, \label{eq:corrsnm}
\end{equation}
where the subindices $n,m$ represent the position in the lattice and $A_n$ are generic operators. The mean value $\left\langle \cdots\right\rangle$ is taken over the ground state of the system before the quantum quench, i.e. the ground state of $H_0$: $\rho_0= \left|\psi_0\right\rangle \left\langle\psi_0\right|$. We work in the thermodynamic limit $L \rightarrow \infty$ which we impose by taking the analytic limit or considering a system of $L=1000$ lattice sites in the case of numerical results. In the limit $t_0\rightarrow\infty$,  correlation functions reach a stationary regime, in which, as in  equilibrium,  they only depend on the time difference $t$: $C_{\pm}^A(t,t_0\rightarrow\infty) =C_{\pm}^A(t)$. This regime is relevant for extracting effective temperatures and is imposed analytically, by using the Riemann-Lebesgue lemma, or numerically, by taking $ t_0=100 $. Within the thermodynamic limit and the stationary regime the linear response function $R_\pm$ and correlator $C_\pm$ of all the operators studied in this paper, show an independence on specific site $n$ and $m$ for periodic boundary conditions; they only depend on the site difference $l = n-m $, $C_{\pm,(n,m)}^A(t,t_0\rightarrow\infty) =C_{\pm,l}^A(t)$. In the case of open b.c, this rule does not apply, but is nearly fulfilled by taking $n$ and $m$ near the center of the lattice.

We shall study the time dependence of $C_\pm$ and $R_\pm$ for several operators in the limits mentioned above, analyzing their dependence with site difference $l$, the initial superlattice potential strength $\Delta$ and initial temperature $T$.

\subsubsection{Local Operators}

\begin{figure}[ht]
\centering
\subfloat{\label{fig:ReCf}} \subfloat{\label{fig:ReRf}}  \subfloat{\label{fig:ImCf}} \subfloat{\label{fig:ImRf}}
				
\includegraphics{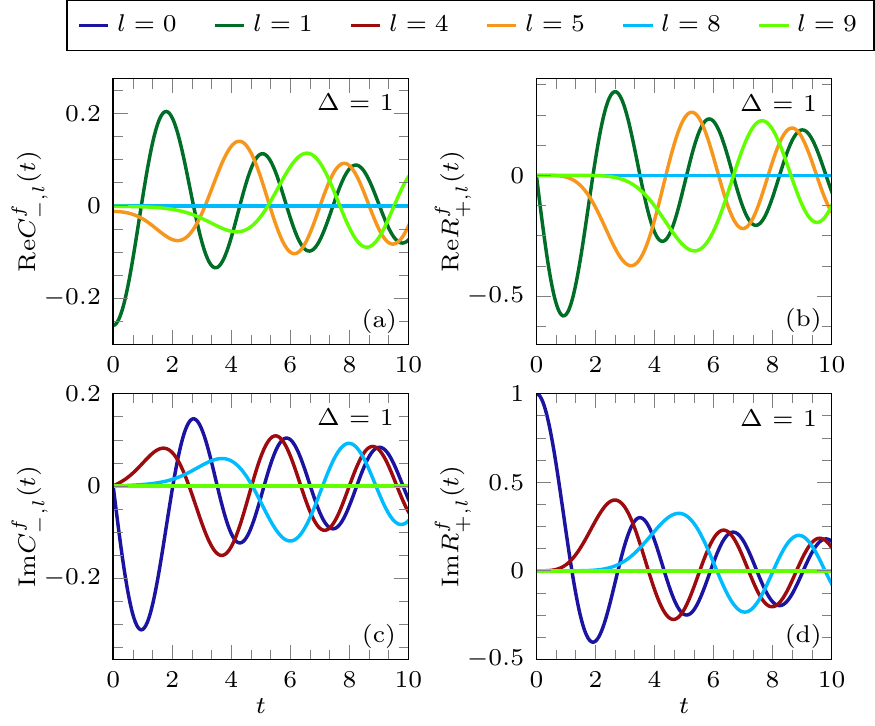}

\caption{Correlation functions for the Fermi operators varying the site difference $l$ with $\Delta =1$. In the panels \protect\subref{fig:ReCf} and \protect\subref{fig:ImCf} we show  the real and imaginary part of $C^f$, respectively, while in the right panels  \protect\subref{fig:ReRf} and \protect\subref{fig:ImRf}  the same information is displayed for the response function $R^f$. }
\label{fig:Rc}
\end{figure}

Let us start by studying the quasiparticle Fermi operator $f_n$  correlation functions. Following the definition \eqref{eq:corrs}, we shall consider
\begin{equation}
C_{\pm,(n,m)}^f(t,t_0)= \langle \left[ f_n(t+t_0),f^\dagger_m(t_0) \right]_\pm \rangle\label{eq:C+c},
\end{equation}
where we shall employ $C_+$ to build the linear response (retarded) function $R_+$. As we mentioned before, in the thermodynamic ($L\rightarrow \infty$) and stationary ($t_0\rightarrow \infty$) limits, these functions have only dependence on $t_0$ and the lattice site difference $l=n-m$. In these regimes, the linear response function $R_+$ results
\begin{equation}
R^f_{+,l}(t) =  \begin{cases}
-t\theta(t)\phantom{}_{1}\tilde{F}_{2}\left(1;\frac{3-l}{2},\frac{3+l}{2};- t^{2}\right) & \text{$l$ odd}\\
\phantom{-}i \theta(t)\phantom{}_{1}\tilde{F}_{2}\left(1;\frac{2-l}{2},\frac{2+l}{2};- t^{2}\right) & \text{$l$ even}\\
\end{cases}
\label{eq:Cc}
\end{equation}
where $ \phantom{}_{1}\tilde{F}_{2}$ represents a generalized (regularized) hypergeometric function. Interestingly, $R^f_{l}$  is independent of $\Delta$  which may lead to the conclusion that in the stationary regime the initial state correlations have been lost. Nevertheless, some information remains as $R^f$ is different for even and odd site difference, which is a consequence of the different translational symmetries of $H$ and $H_0$. On the other hand, the antisymmetric correlator $C^f_-$ in the stationary regime,
\begin{equation}
C_{-,l}^f(t) =  \intop_{-\pi/2}^{\pi/2} \frac{dk}{4 \pi} \cos (k l) \frac{\omega_k \left(e^{-i \omega_k t}-e^{i\pi l}e^{i \omega_k t}\right) }{\sqrt{\omega_k^2+ \Delta^2}}
\label{eq:Rc}
\end{equation}
 does depend on the supperlatice potential. In Fig. \ref{fig:Rc} we plot the real and imaginary parts of the response $R_+$ and antisymmetric correlator $C_-$ for different values of site difference $l$.
We observe that both functions are real or pure imaginary for odd $l$ or even $l$, respectively. Also, we notice the presence of the so-called light-cone effect~\cite{calabrese06_quench_CFT}, in which the functions are expected to be constant up to a time $t_e = l/v_e$ ($l/2$ in this case) where $v_e$ is the quasiparticle (excitation) velocity. On the other hand, the change in $\Delta$ reduces the amplitude of $C_-^f$. Moreover, for large $\Delta$,
\begin{equation}
C_{-,l}^f(t)  \approx    \begin{cases}
-\frac{\sqrt{\pi}\phantom{}_{2}\tilde{F}_{3}\left(1,\frac{3}{2};\frac{1}{2},\frac{3-l}{2},\frac{3+l}{2};- t^{2}\right)}{2\Delta} & \text{$l$ odd}\\
-\frac{i t\phantom{}_{1}\tilde{F}_{2}\left(1;\frac{2-l}{2},\frac{2+l}{2};- t^{2}\right)}{\Delta} & \text{$l$ even},
\end{cases}
\label{eq:CcD}
\end{equation}
while the long time behavior is well represented by
\begin{equation}
C_{-,l}^f(t)  \approx   \frac{\alpha}{\sqrt{t}} I_l \cos \left( 2 t +  \phi_l \right) \label{eq:Rct},
\end{equation}
where $\alpha = \alpha (\Delta)$, $\phi_l$ a phase that depends on the site difference and $I_l=i$ for even $l$ and $I_l=1$ in other case. The decay rate is universal ($t^{-1/2}$), clearly independent from $l$ or $\Delta$.  Both $R$ and $C$ in the stationary regime, show the same decay rate as the density and one time $ \langle f^\dagger_{-l}(t) f_l(t)  \rangle$ correlation functions.  As we shall see, the rather simple structure of the Fermi operator correlation functions will allow us to extract a simple expression for the effective temperature, which coincides with the one expected in the GGE.

\begin{figure}[ht]
\centering
\subfloat{\label{fig:Cnl}}
\subfloat{\label{fig:Rnl}}
\subfloat{\label{fig:CnD}}
\subfloat{\label{fig:RnD}}

\includegraphics{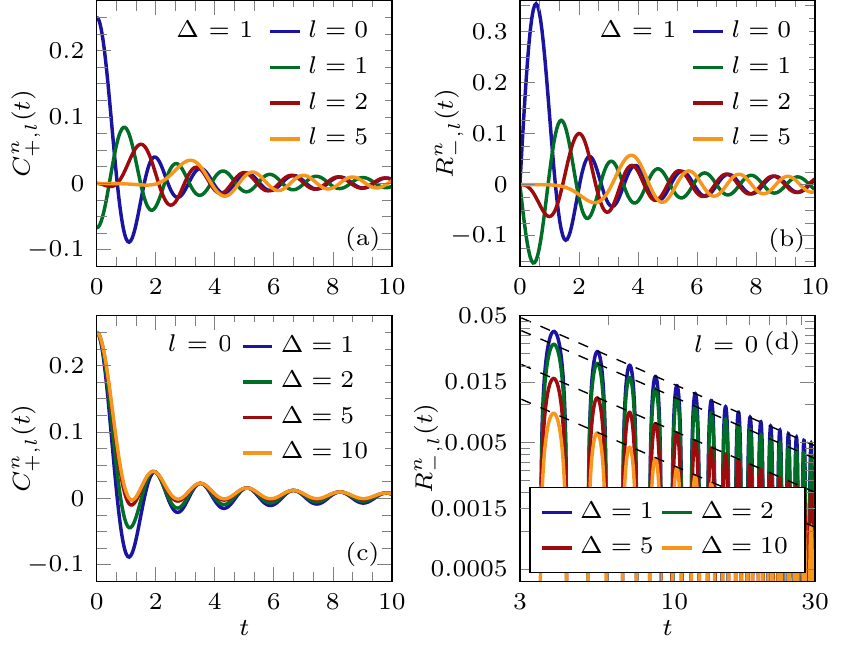}

\caption{Density-density two-time correlation functions. Figs. \protect\subref{fig:Cnl} and \protect\subref{fig:Rnl} show the site difference dependence for $\Delta =1 $ ($C_+^n$ in \protect\subref{fig:Cnl}, $R_-^n$ in \protect\subref{fig:Rnl}). The change in $\Delta$ for $C_+^n$ \protect\subref{fig:CnD} and $R_-^n$ \protect\subref{fig:RnD} with $l=0$. In the double logarithmic scale plot \protect\subref{fig:RnD} the dashed lines represent a $t^{-1}$ decay, compatible with both correlators. }
\label{fig:dens-dens}
\end{figure}

At this point one wonders whether the properties observed above are unique of the quasiparticle correlations or manifest in other type of correlation functions. For instance, we shall consider the case of the density-density correlator,
\begin{equation}
C_{+,l}^n(t,t_0)= \langle \left[ n_n(t+t_0), n_m(t_0) \right]_\pm \rangle.
\end{equation}
 As $n_i(t)$ is a bosonic operator, we study the usual correlation functions $R_-$ and $C_+$. Fig. \ref{fig:dens-dens} shows the $\Delta$ and lattice site difference $l$ dependence of these functions in the stationary regime. Both functions show a $t^{-1}$ universal decay,
\begin{align}
C_{+,l}^n(t) \approx & \frac{\alpha_c}{t} (\beta + \sin 4t)\\
R_{-,l}^n(t) \approx & \frac{\alpha_r}{t} \cos 4t
\end{align}
for  the $t\gg 1$ regime, which is also shown by the out of equilibrium one time density correlation  $\left \langle n_{l}(t) n_j(t) \right \rangle$. The light-cone effect is also present. As in the previous correlators an increase in the initial superlattice potential intensity decreases the correlation functions amplitude. In the large $\Delta$ limit, both functions can be written as the product of hypergeometric functions:
\begin{align}
C_{+,l}^n(t) \approx & (-1)^l (P_l^2(t)+Z_l^2(t))\label{eq:Cnt} \\
R_{-,l}^n(t) \approx & -4 i (-1)^l \theta(t)P_l(t)Z_l(t) \label{eq:Rnt}
\end{align}
where $P_l$ represents $\phantom{}_{1}\tilde{F}_{2}\left(1;1-\frac{l}{2},1+\frac{l}{2};- t^{2}\right)/2$ for even $l$ and  $-\phantom{}_{1}\tilde{F}_{2}\left(1;\frac{3-l}{2},\frac{3+l}{2};- t^{2}\right)i t/2$ for odd $l$ whereas $Z_l$ is  $-\phantom{}_{1}\tilde{F}_{2}\left(2;2-\frac{l}{2},2+\frac{l}{2};- t^{2}\right)it/\Delta$ when $l$ is even and $-\phantom{}_{2}\tilde{F}_{3}\left(1,\frac{3}{2};\frac{1}{2},\frac{3-l}{2},\frac{3+l}{2};- t^{2}\right)\sqrt{\pi}/2\Delta$ in the other case.

\subsubsection{Non-local operators}

\begin{figure}[ht]
\vspace*{-12pt}
\subfloat{\label{fig:nolocC}}
\subfloat{\label{fig:nolocR}}

\includegraphics{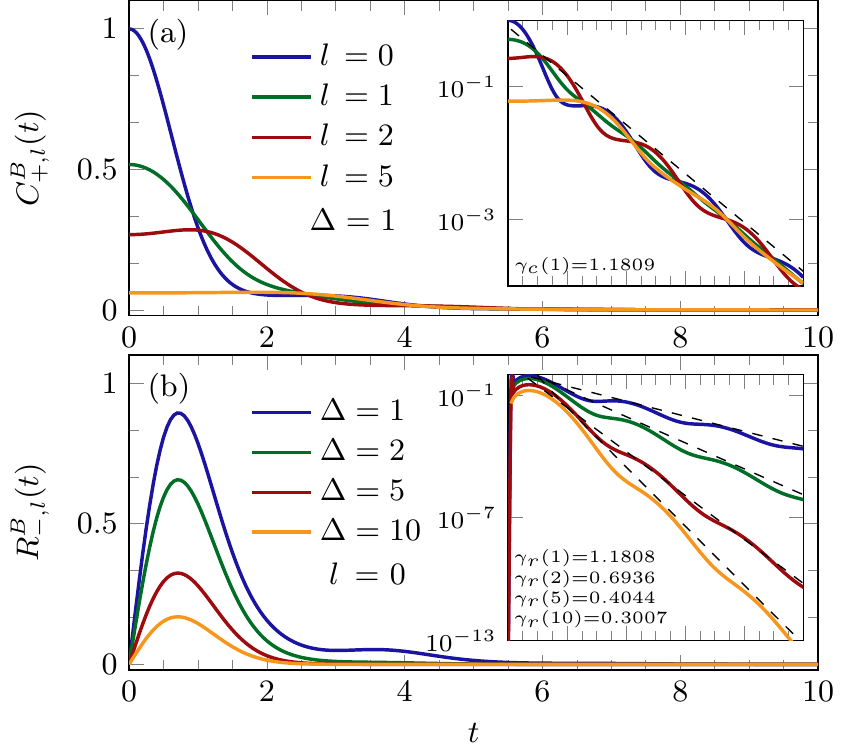}
\caption{Non-local two-time correlation functions. Figs. \protect\subref{fig:nolocC} and \protect\subref{fig:nolocR} show the site difference dependence for $C^B_+$ and the $\Delta$ dependence in $R^B_-$, respectively. In the insets the same information in semi logarithmic axis along with exponential decays. The insets also present the exponential decay constants $\gamma_c(\Delta)$ and $\gamma_r(\Delta)$ respectively. }
\label{fig:noloc}
\end{figure}

The last set of operators we shall consider are the hard-core bosons creation and annihilation non-local operators  $b_n$ and  $b_m^\dagger$ written in terms of the local operators as
\begin{equation}
 b_{n}=\prod_{j=1}^{m-1}(1-2f_{j}^{\dagger}f_{j}) f_{m},\quad b_{m}^{\dagger}=f_{m}^{\dagger}\prod_{j=1}^{m-1}(1-2f_{j}^{\dagger}f_{j}). \label{eq:jw}
\end{equation}
Non-local two-time correlations have been already studied in Refs.  \onlinecite{rossini10_quantum_ising_quench,foini11_quench_FDT,foini12_quench_FDT_long} for the quantum Ising model in a transverse magnetic field. In these papers the computation $\langle \sigma^x_n(t+t_0) \sigma^x_m(t_0)\rangle$ involves calculating the four-spin correlation function done by means of a $2L\times 2L$ T\"{o}plitz determinant. The two-spin correlator is then recovered by taking the thermodynamic limit and making use of the cluster property. For our model,  the fermionic Hamiltonian (equation \eqref{eq:H0}) does not contain anomalous terms and therefore we can make use of a simpler straightforward approach. We start by defining the hermitian combination $B_i=b_i+b_i^\dagger$  and considering the two-point correlation functions $C_\pm^B$,
\begin{equation}
C_\pm^B(t,t_0)= \langle \left[ B_n(t+t_0), B_m(t_0) \right]_\pm \rangle \label{eq:noloc2t}
\end{equation}
from which we can calculate the response function. We observe that only one of the two terms in equation \eqref{eq:noloc2t} is needed, as $C_+^B=\Real \langle B_n B_m \rangle$ and $C_-^B=i\Imag \langle B_n B_m \rangle$. Using the definition, we obtain
\begin{equation}
\langle  B_n(t) B_m(t')  \rangle = \langle  b_n(t) b_m^\dagger(t')  \rangle+ \langle  b_n^\dagger(t) b_m(t')  \rangle
\label{eq:nlB}
\end{equation}
since the remaining terms vanish. The first term in equation \eqref{eq:nlB} can be computed by extending the approach presented  in Ref. \onlinecite{rigol05_groundstate_hcbosons} for different times. We can write
\begin{equation}
\left\langle \Psi\right|b_n(t)b_m^\dagger(t')\left| \Psi \right\rangle= \left\langle \Psi(t)\right|b_n e^{-i Ht}e^{i Ht'} b_m^\dagger\left| \Psi(t') \right\rangle \label{eq:noloc1}
\end{equation}
where $b_m^\dagger (b_n)$ can be mapped to fermions by the equation \eqref{eq:jw} and $\Psi(t)$ is the time evolved ground state:
\begin{align}
\left| \Psi(t') \right\rangle &= \prod_{\nu=1}^N  e^{-iHt'} c_\nu^\dagger |0\rangle= \prod_{\nu=1}^N  e^{-iHt'} c_\nu^\dagger e^{iHt'} e^{-iHt'}  |0\rangle \nonumber \\
&= \prod_{\nu=1}^N \sum_{j=1}^L f_j^\dagger \varphi_{\nu}(j,t')\left| 0\right\rangle
\end{align}
where $c^\dagger_\nu$ are the operators that render $H_0$ diagonal and  $\varphi_{\nu}(j,t')$ the time dependent eigenfunctions of $H_0$. Then
\begin{equation*}
 b^\dagger_m \left| \Psi(t') \right\rangle = f_{m}^{\dagger}\prod_{j=1}^{m-1}(1-2f_{j}^{\dagger}f_{j}) \prod_{\nu=1}^N \sum_{j=1}^L f_j^\dagger \varphi_{\nu}(j,t')\left| 0\right\rangle.
\end{equation*}
Then we define a $L\times N$ matrix $P(t')$ with elements  $\varphi_{\nu}(j,t')$. Then the action of $b_m^\dagger$ on $\left| \Psi(t') \right\rangle$ amounts to change the signs of elements $P_{j\nu}$ with $j\leq m-1$ and the further creation of a particle at site $m$ implies the addition of a column to $P$ with elements $P_{i,N+1}=\delta_{im}$. Thus, we can write
\begin{align}
e^{i Ht'} b_m^\dagger\left| \Psi(t') \right\rangle&= \prod_{\nu=1}^{N+1} \sum_{j=1}^L f_j^\dagger(t') P'_{j\nu}(t')\left| 0\right\rangle\\
&= \prod_{\nu=1}^{N+1} \sum_{i=1}^L f_i^\dagger Q_{i\nu}(t') \left| 0\right\rangle
\end{align}
where $P'$ is obtained by changing the required signs and adding the new column, and $Q(t')= e^{iht'}P'(t)$ is again a  $L\times N$ matrix, where $h$ is the matrix representation of the Hamiltonian $H$. Hence, we can rewrite equation \eqref{eq:noloc1} as
\begin{align}
\left\langle \Psi\right|b_n(t)b_m^\dagger(t')\left| \Psi \right\rangle&= \det Q^\dagger(t) Q(t')\\
&= \det P'^\dagger_n  (t)e^{-ih(t-t')} P'_m(t') \label{eq:nl1}.
\end{align}
The second term in the correlator \eqref{eq:nlB} is more involved since we can no longer create a new column in $P$ as the fermionic creation and destruction operators are permuted with respect to the ground state operators,
\begin{align}
\langle  b_n^\dagger(t) b_m(t')  \rangle =&  \prod_{\mu,\nu=1}^N\sum_{j,l=1}^L \varphi^\ast_{\mu}(l,t) \varphi_{\nu}(j,t') \times \nonumber  \\
&\left\langle 0 \right| f_l \cdots f_{n}^{\dagger} e^{-i H(t-t')} f_{m}\cdots f_j^\dagger \left| 0\right\rangle
\end{align}
We circumvent this issue by employing the following property: Calling $\tau = t-t'$:
\begin{align}
f_n^\dagger e^{-iH\tau}f_m=&  e^{-iH\tau} f_n^\dagger(\tau) f_m = e^{-iH\tau}\sum_{j=1}^N f_j^\dagger (e^{ih \tau})_{jn} f_m \nonumber \\
=- e^{-iH\tau}& f_m e^{iH\tau}f_n^\dagger e^{-iH\tau} +e^{-iH\tau}(e^{ih \tau})_{mn}\label{eq:nl2}
\end{align}
Then $\langle  b_n^\dagger(t) b_m(t')  \rangle$   can be written as
\begin{align}
\langle  b_n^\dagger(t) b_m(t')  \rangle =& \det P^{\dagger}_n (t) e^{-ih (t-t')} P_m(t')\nonumber \\& \quad - \det O_n^{m\dagger}(t,t') O_m^{n}(t',t)
\end{align}
where $P_m$ is $P'_m$ with no additional column and $O_m^n$ is a $L \times N+1$ matrix defined by
\begin{equation}
O_m^{n}(t',t)=
\begin{cases}
(e^{iht'}P^m(t'))_{j \nu} & \mbox{for } \nu = 1, \ldots  ,N \\
(e^{iht})_{j n} & \mbox{for } \nu = N+1
\end{cases}
\end{equation}
for $j = 1, \ldots , L $. We can recover $\langle B_n (t+t_0)B_m(t_0)\rangle $ by adding expressions \eqref{eq:nl1} and \eqref{eq:nl2} and taking $t'\rightarrow t_0$ and $t\rightarrow t+t_0$. Thus, for our model, this approach reduces the computation of non-local correlations to the evaluation of $(N+1)\times (N+1)$ matrix determinants, instead of determinants of $2L\times 2L$ T\"{o}plitz matrices. 
 We compute the non-local correlation function using a system with 1000 lattice sites with open-boundary conditions, half-filled ($N=L/2$) and taking $t_0 = 100$ as the stationary limit. In Fig. \ref{fig:noloc} we show the results obtained for $C_+$ (\ref{fig:nolocC}) and linear response function (\ref{fig:nolocR}).
These functions present an exponential decay whose rate depends on the initial superlattice potential $\Delta$, and is independent of the lattice difference (shown in the insets of Figs. \ref{fig:nolocC} and \ref{fig:nolocR}). The long time behavior is well fitted by
\begin{align}
C_{+,l}^{B}(t)=& \alpha_c e^{- \gamma_c t} \left( \beta_c + \frac{\sin (2t+\phi)}{\sqrt{t}} \right)\\
R_{-,l}^{B}(t)=& \alpha_r e^{- \gamma_r t} \left( \beta_r + \frac{\sin (2t+\phi)}{\sqrt{t}} \right)
\end{align}
i.e, damped oscillations modulated by an exponential decay dictated by $\gamma_i = \gamma_i (\Delta)$.

\subsubsection{Initial state at finite temperature}
\label{sec:timetemp}
\begin{figure}[ht]
\vspace*{-12pt}
\subfloat{\label{fig:ImCcparT}} \subfloat{\label{fig:ReCcimparT}} \subfloat{\label{fig:CnT}}\subfloat{\label{fig:RnT}}

\includegraphics{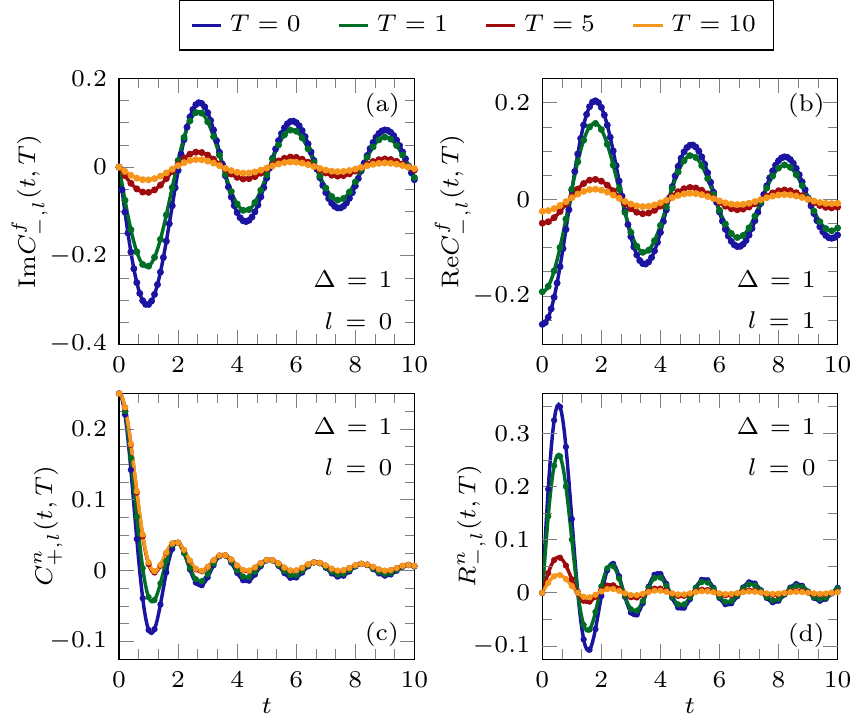}
\caption{Initial temperature dependence of the two-time correlators. Figs.  \protect\subref{fig:ImCcparT} and \protect\subref{fig:ReCcimparT}  show the $C_-$ correlator for Fermi operators with different site difference, while \protect\subref{fig:CnT} and \protect\subref{fig:RnT} are $C_+$ and $R_-$ for density operators. In both cases $\Delta=1$ and $l=0$. The solid lines represent periodic boundary conditions while the dots are open boundary conditions correlators.}
\label{fig:timetemps}
\end{figure}

We extend our analysis to the case in which the initial state is a thermal state with temperature $T$,  described by $\rho_0 = \exp(-H_0/T)/Z$, which involves working in the grand canonical ensemble (GCE).   This raises a new problem  as the border terms $f_L^\dagger f_{L+1}$ are treated by imposing (anti-)periodic boundary conditions which depend on the number of particles $N$ in the system, and $N$ is not fixed in the GCE. One possible workaround could be to calculate the correlations using open boundary conditions, but this approach complicates the analytical results. We address this issue by keeping the simplicity of analytically calculated periodic boundary conditions correlators and checking the relevance of the border terms comparing these results with the ones obtained by solving the problem numerically with open-boundary conditions (shown as dots in Fig. \ref{fig:timetemps}). We checked the independence of the boundary conditions for the correlators in the zero temperature case far from the lattice borders.

Following the zero temperature analysis done before, we start by studying the Fermi operator correlators. We compute $C_{\pm,l}^f(t,t_0)= \langle \left[ f_n(t+t_0),f^\dagger_m(t_0) \right]_\pm \rangle$ where $\left\langle \cdots\right\rangle $ now represents $\Tr [\rho_0 \cdots]$. In the thermodynamic limit and stationary regime the $R_+$ correlator is the same as in the  zero temperature case (equation \eqref{eq:Cc}), i.e. it has neither $\Delta$ nor initial temperature dependence. The differences between this result and the one obtained numerically with open boundary conditions are negligible. The temperature $T$ and superlattice potential $\Delta$ dependences are only contained in the linear response function
\begin{equation}
 C_{-,l}^f(t,T)=  \intop_{-\pi/2}^{\pi/2} dk\; \mathcal{I}\, \tanh(\sqrt{\omega_k^2+ \Delta^2}/2T),
\end{equation}
where $\mathcal{I}$ represents the integrand in equation \eqref{eq:Rc}. In Figs. \ref{fig:ImCcparT} and \ref{fig:ReCcimparT} we show this function (solid lines) and the numerical calculations (dots) varying the reservoir temperature $T$. The agreement of both calculations, periodic and open boundary, shows that the border terms are not significant.
We notice that the limit $T \rightarrow 0$ is well defined as we recover the zero temperature result. Varying the initial temperature has a similar behavior in $C_-$ as changing the supperlattice potential strength $\Delta$. Moreover, the large $T$ limit as in the Fermi case is identical to  $\Delta \gg 1$ regime (equation \eqref{eq:CcD})  taking $\Delta = 2T$, while the strong insulator limit is the same as in the $T=0$ case. Furthermore, for large time difference ($t\gg 1$)  it has the same behavior as in zero temperature, shown in equation \eqref{eq:Rct}, with $\alpha = \alpha(\Delta,T)$.


The analysis of density-density correlators with an initial thermal state, shown in Figs. \ref{fig:CnT} and \ref{fig:RnT}, shows similar features than the Fermi correlators. The effect of rising $T$ is similar to the one produced by increasing $\Delta$ and the high temperatures limits is well described by equations \eqref{eq:Cnt} and \eqref{eq:Rnt} taking $\Delta=2T$.  As in the Fermi case, open (dots) and periodic boundary (lines) conditions correlators coincide, showing that the border terms do not play an important role in the studied correlations.

\subsection{Frequency dependence}
\label{sec:corrfrec}
In this section we analyze the frequency dependence of the correlation calculated in section \ref{sec:corrtime}. More specifically, we study the Fourier transform of the linear response function imaginary part and the (anti-)symmetric correlator in the stationary and thermodynamic limits, both of the functions related by the fluctuation-dissipation theorem. Following the order established in section \ref{sec:corrtime}, we start by analyzing the simpler Fermi correlations, whose linear response function imaginary part in the frequency space is
\begin{equation}
\mathrm{Im}R^f_{+,l}(\omega) = \theta \left(1- \omega^2/4\right) \frac{e^{i \pi l}T_l(\omega/2)}{\sqrt{1-\omega^2/4}}, \label{eq:Cfom}
\end{equation}
where  $T_n(x)$ are the Chebyshev polynomials of the first kind and degree  $n$. The higher contribution to $R_+$ comes from frequencies from the bands' edge ($\omega \approx\pm 2$), while the $T_n$ polynomials mostly modify the center of the band as the site difference $l$ increases. Furthermore, the antisymmetric correlator is
\begin{equation}
C_{-,l}^{f} (\omega) = \mathrm{Im} R_{+,l}^f (\omega) \frac{\omega}{\sqrt{\omega^2+ \Delta^2}},
\end{equation}
which shows the same bandwidth and functional dependence in $l$. The main effect of increasing the supperlatice potential strength is to reduce the contribution of the frequencies in the center of the band to $C_-$. Since $\mathrm{Im}R_+$ can be factorized from $C_-$, the effective temperature can be easily extracted (see equation  \eqref{eq:Teffc}). On the other hand, when the system is in contact with a thermal reservoir the temperature dependence appears in $C_-^f$ through a multiplicative factor,
\begin{equation}
C_{-,l}^{f} (\omega,T) = C_{-,l}^{f} (\omega) \tanh \frac{\sqrt{\omega^2+\Delta^2}}{2T}.
\end{equation}
Even though it clearly modifies the response function, the main consequence of rising the temperature of the initial reservoir is similar to the one produced by  increasing $\Delta$:  decreasing the contribution of the low frequency modes in the correlation function 
as $T \rightarrow \infty$. As expected from the results shown in section \ref{sec:timetemp}, the linear response function is independent of $T$, coinciding with equation \eqref{eq:Cfom}.

\begin{figure}[ht]
\vspace*{-12pt}
\subfloat{\label{fig:Comega}} \subfloat{\label{fig:ImromegaT}} \subfloat{\label{fig:Cnolocomega}}\subfloat{\label{fig:Imrnolocomega}}

\includegraphics{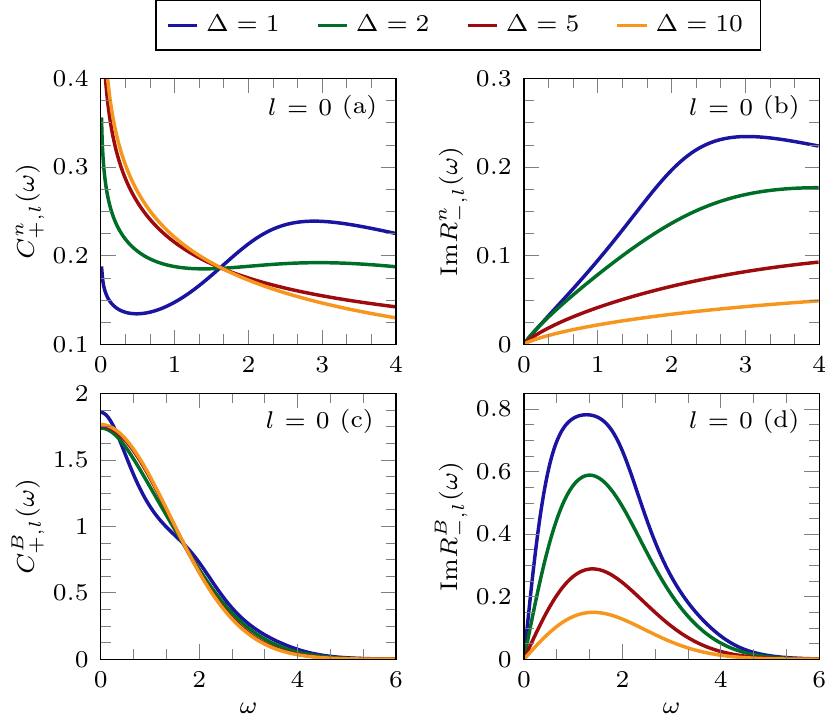}
\caption{Density and non-local correlators in frequency space. Panel \protect\subref{fig:Comega}  and \protect\subref{fig:ImromegaT} show the density symmetric correlator and  linear response imaginary part for different $\Delta$ values. Panels \protect\subref{fig:Cnolocomega} and \protect\subref{fig:Imrnolocomega} present the same functions for non-local operators. In all the cases $l=0$.}
\label{fig:frecs}
\end{figure}

Next, we study the frequency dependence of the density and non-local correlations in the thermodynamic limit and stationary regime, by performing a discrete Fourier transform over the time-dependent correlators in $t\in[0,100]$ with a time interval $\tau=0.25$. In Fig. \ref{fig:frecs} we plot these functions, only showing the positive frequency sector as both functions have definite parity ($C_+$ is even and $\textrm{Im}R_-$ is odd). Both density correlators (Figs. \ref{fig:Comega} and \ref{fig:ImromegaT}) present a contribution from frequencies between $-4 \leq\omega \leq 4$. For small values of $\Delta$ the contribution of higher frequencies to $C_+^n$ is important, but as the initial potential increases the lower frequency modes become more relevant.  In the case of $\textrm{Im}R_-^n$, the amplitude seems to be inversely proportional to $\Delta$, decreasing the contribution of all frequency modes for higher potential values. Finally, the non-local correlators present a different panorama, as both functions amplitude decrease as the frequency increases. Analyzing the variation with $\Delta$, we  notice that the symmetric correlator remains almost unchanged, only becomes smoother with this change. The linear response imaginary part presents a peak around $\omega\sim 1.3$, which reduces its amplitude and shifts to higher frequencies as the initial superlattice potential rises. The frequency-dependent correlators obtained in this section shall be employed in the calculation of effective temperature, depicted in section \ref{sec:efftemp}.

\section{Effective temperatures from FDRs}
\label{sec:fdr}
In this section we compute the effective temperatures from the correlators studied in section \ref{sec:corr}, analyzing both zero and finite temperature initial states. Let us start by stating some generalities of the fluctuation-dissipation theorem (FDT).
For typical observables having bosonic properties, the correlation function $C^-$ is used to construct the retarded function $R_-^{AB}(t,t_0)=2i\theta(t)C^{AB}_-(t,t_0)$, while, in the case of Fermi operators which do not commute, the retarded function is defined employing the commutator, $R_+^{AB}(t,t_0)=2i\theta(t)C^{AB}_+(t,t_0)$. The FDT relates the functions $R_\pm^{AB}$ and $C_\mp^{AB}$ in equilibrium at inverse temperature $\beta$. In the frequency domain, where
\begin{equation}
R_\pm^{AB}(\omega)=\int_{-\infty}^{\infty} dt e^{i\omega t} R_\pm^{AB}(t),
\end{equation}
it takes the form
\begin{equation}
\Im R_\pm^{AB}(\omega)=\left[\tanh\frac{\omega\beta}{2}\right]^{\mp 1}C_\mp^{AB}(\omega).
\end{equation}

Before obtaining specific results for effective temperatures from FDT for this model, let us state a general result valid for quasi-free systems whose static correlations relax to the GGE. In this case dynamic correlations of local operators are also asymptotically described by the GGE~\cite{essler12_dynamical_gge}. By using a spectral decomposition in terms of eigenstates of the Hamiltonian one can show that a basic form of the FDT holds out of equilibrium for long times~\cite{essler12_dynamical_gge}:
\begin{equation}
-\frac{1}{\pi}\Im\chi_{\mathcal{A}\mathcal{B}}\left(\omega\right)=S_{\mathcal{A}\mathcal{B}}(\omega)-S_{\mathcal{B}\mathcal{A}}(-\omega).
\end{equation}
However, differently from the usual FDT for systems in thermodynamic equilibrium, the negative $S_{\mathcal{A}\mathcal{B}}(\omega)$ and positive $S_{\mathcal{B}\mathcal{A}}(-\omega)$ parts of the spectral function in general are not simply related by $S_{\mathcal{B}\mathcal{A}}(-\omega)=e^{-\beta\omega}S_{\mathcal{A}\mathcal{B}}(\omega)$, where $\beta$ is the inverse temperature. We will show that after relaxation from a quantum quench it is possible to establish an analogous relation for correlations of quasiparticle creation and destruction operators.

Consider a general bilinear Hamiltonian $H_b=\sum_{i,j} f^\dagger_i h_{ij} f_j$ where $f^\dagger_i$ and $f^\dagger_j$ are destruction and creation fermionic operators and $h$ a symmetric matrix. $H_b$ is diagonalized by a canonical transformation $f_j=\sum_\nu\mathcal{U}_{j,\nu}f_\nu$, $H_b=\sum_\nu\varepsilon_\nu f^\dagger_\nu f_\nu$ where $\varepsilon_\nu$ is the dispersion relation. Consider the correlation function for the Fermi field
\begin{align}
C^{ij}(t,t_0)=&\langle f_i(t+t_0)f_j^\dagger(t_0)\rangle\\ =&\sum_{\mu\nu}\mathcal{U}^\ast_{i\mu}\mathcal{U}_{j\nu} e^{-i\varepsilon_\nu(t+t_0)}e^{i\varepsilon_\mu t_0}\langle f_\mu f^\dagger_\nu\rangle\label{eq:correlator}
\end{align}
where $\langle\ldots\rangle$ is the initial state. Even though the correlator $\langle f_\mu f^\dagger_\nu\rangle$ is not diagonal for initial states that are not translation invariant, for rather standard conditions the non diagonal contributions decay rapidly and vanish in the thermodynamic limit\cite{cazalilla12_thermalization_correlations,barthel08_quench} which constitutes the way by which dephasing takes place. In the specific model we are analyzing, the eigenmode correlator is not diagonal in momentum space, but the only contribution outside the diagonal is the correlation between modes at $k$ and $k+\pi$, $\langle f_k f^\dagger_{k+\pi}\rangle =\Delta/E_k$. In the thermodynamic limit these terms yield a smooth function of $k$ and therefore by application of the Riemann-Lebesgue theorem do not contribute to Eq. (\ref{eq:correlator}):
\begin{equation}
\lim_{t_0\to +\infty}C^{ij}(t,t_0)=\sum_{\mu}\mathcal{U}^\ast_{i\mu}\mathcal{U}_{j\mu} e^{-i\varepsilon_\mu t}\left[1-N_0(\varepsilon_\mu)\right]
\end{equation}
where $N_0(\varepsilon_\nu)=1/[e^{\lambda(\varepsilon_\nu)}+1]$ are the mode occupations in the initial state. From this correlator we can construct the response and the correlation function, which in frequency space read
\begin{align}
\Im R_+(\omega)=&\pi\sum_{\mu}\mathcal{U}^\ast_{i\mu}\mathcal{U}_{j\mu} \delta(\omega-\varepsilon_\mu)\\
C_-(\omega)=&\pi\sum_{\mu}\mathcal{U}^\ast_{i\mu}\mathcal{U}_{j\mu} \delta(\omega-\varepsilon_\mu)[1-2N_0(\varepsilon_\mu)]
\end{align}
Therefore, both functions are related as
\begin{equation}
\Im R_{+}(\omega)= \left[\tanh\frac{\lambda(\omega)}{2}\right]^{-1} C_-(\omega),
\end{equation}
and therefore we have a frequency-dependent effective temperature $1/ T_\mathrm{eff}(\omega)=\lambda(\omega)/\omega$. We notice that this result is generic for initial states and quenches to quasi-free models for which the long-times regime is captured by the GGE.

\subsection{Effective temperatures for local and non-local operators}
\label{sec:efftemp}

After obtaining this general result, we wish to explore the effective temperatures extracted from the correlators calculated for our model shown in section \ref{sec:corr}.  In general, these can be written as
\begin{equation}
T^{AB}_{\mathrm{eff}}(\omega)=\frac{\omega}{2}\mathrm{arctanh}^{-1}\left[\left(\frac{\Im R_\pm^{AB}(\omega)}{C_\mp^{AB}(\omega)}\right)^{\mp 1}\right].
\end{equation}
For out of equilibrium systems these temperatures usually depend on frequency and the operators studied. However, if the system achieves a thermal state after long times, all of the $T_\mathrm{eff}$ should be equal and frequency independent, at least for a value of $t_0$ large enough.
\begin{figure}[ht]
\centering
\subfloat{\label{fig:Teffn}}\subfloat{\label{fig:TeffB}}\subfloat{\label{fig:TeffnT}}
\includegraphics{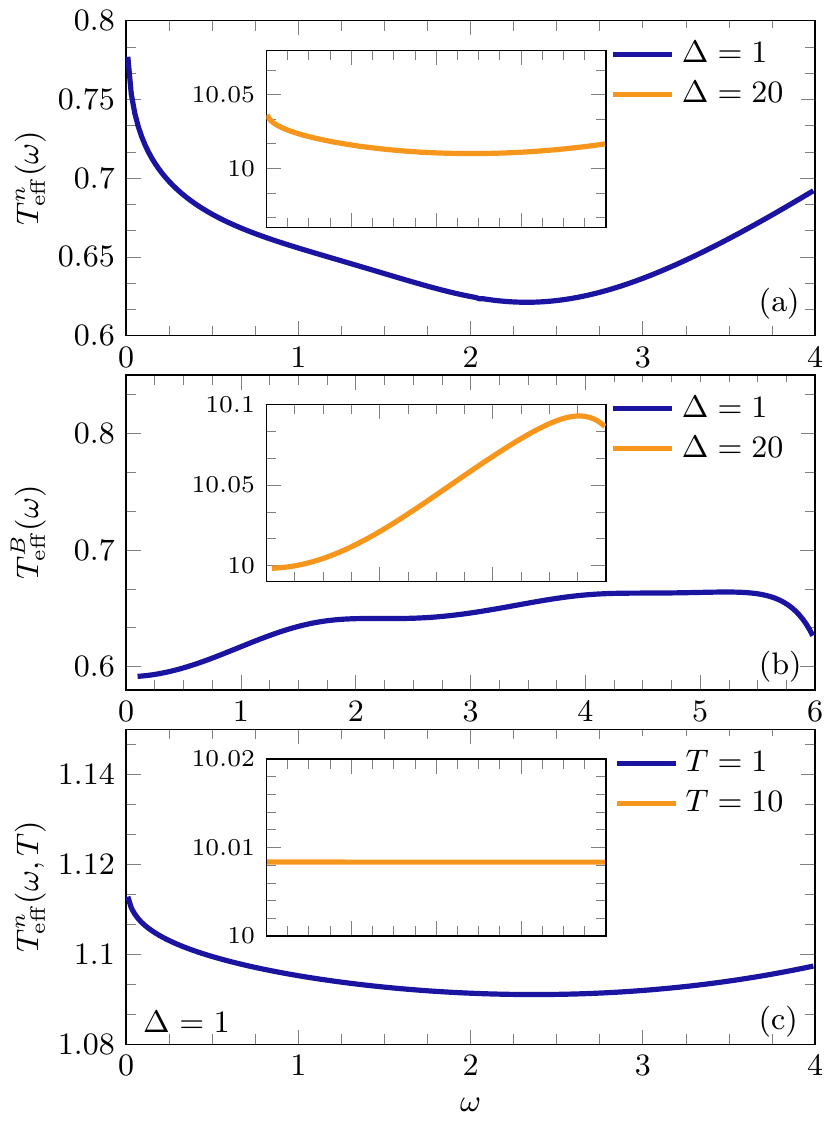}
\caption{Effective temperatures for the different autocorrelation functions: density operators ($T_{\mathrm{eff}}^n(\omega)$) in Fig. \protect\subref{fig:Teffn}, non-local operators ($T_{\mathrm{eff}}^B(\omega)$) in Fig. \protect\subref{fig:TeffB} and density operators ($T_{\mathrm{eff}}^n({\omega,T)}$) for an initial thermal state in Fig. \protect\subref{fig:TeffnT}.}
\label{fig:Teff}
\end{figure}

Let us start with the Fermi operators correlations, whose  effective temperature $T^f_{\mathrm{eff}}$ can be calculated analytically, being
\begin{equation}
T^f_{\mathrm{eff}}(\omega) = \frac{\omega}{2}\mathrm{arctanh}^{-1}\left[\frac{\omega}{\sqrt{\omega^2 + \Delta^2}}\right] \label{eq:Teffc}.
\end{equation}
Thus, we obtain a frequency dependent effective temperature that is is independent of the site difference, even though the correlation functions depend on this difference. Nevertheless, one can check the fidelity of $T_\mathrm{eff}^f$:  by reducing the size of the quench by taking $\Delta \rightarrow 0$, $T_\mathrm{eff}^f\rightarrow 0$ and equilibrium is recovered.  As we expected from the general result above, $T_\mathrm{eff}^f$ coincides with the  temperature calculated in the GGE ($T_\mathrm{eff}^\mathrm{GGE}$)  and therefore is  $T^f_\mathrm{eff} \approx \Delta/2$ in the $\Delta \gg 1$ regime. 

At this point the relevant question is whether these characteristics are shared by the effective temperatures that correspond to other observables. In Figs. \ref{fig:Teffn} and \ref{fig:TeffB} we show the temperatures obtained for the autocorrelation functions ($l=0$) of density and  non-local operators, respectively. As one could expect, they do not share the same frequency dependence and are different from $T_\mathrm{eff}^f$. However,  as  $\Delta$ increases,  the effective temperature from density correlations smooths out and reduce its amplitude approaching the value $\Delta/2$ predicted by the GGE temperature, as is shown in the inset of Fig. \ref{fig:Teffn}. Although in this regime the system seems to approach a standard Gibbs ensemble with temperature $T=\Delta/2$, the remaining frequency dependence, as in the case of $T_\mathrm{eff}^f$, discards thermalization. In the non local case (Fig.  \protect\subref{fig:TeffB}) the effective temperature seems to approach $\Delta/2$ for large values of $\Delta$. However its deviations from this value at intermediate frequencies are larger than in the local case, and do not vanish in the limit $\Delta\to\infty$. 

When the system is connected with a thermal reservoir before the quench, the properties of the effective temperatures  are quite similar to the ones above. For the Fermi operators, the additional temperature dependence in $T_\mathrm{eff}^f(\omega,T)$ is given by an extra factor in the argument of the hyperbolic arctangent,
\begin{equation}
T^f_{\mathrm{eff}}(\omega,T) = \frac{\omega}{2}\mathrm{arctanh}^{-1}\left[\frac{\omega  \tanh \left( \frac{\sqrt{\omega^2 + \Delta^2}}{2T}\right)}{\sqrt{\omega^2 + \Delta^2}} \right].
\end{equation}
As $T_\mathrm{eff}^f(\omega)$, it shows an independence on the site difference $l$. It also presents a well defined ``equilibrium'' limit approaching $T$ as  $\Delta \rightarrow 0$, while in the $\Delta \gg 1$ regime follows the GGE temperature. As expected by the results in section \ref{sec:corr}, the high temperature regime is $T_\mathrm{eff}^f(\omega,T) \approx T$, but as a residual frequency dependence remains, a thermal state is not reached in this regime. The density-density autocorrelation function, shown in Fig. \ref{fig:TeffnT}, presents a similar panorama. Its frequency dependence is different from the correlators above, although as $\Delta$ or $T$ rises its value approaches $\Delta/2$ or $T$, respectively. Comparing Figs. \ref{fig:Teffn} and \ref{fig:TeffnT}, it seems that one can reach a state similar to a standard Gibbs state faster by increasing the reservoir temperature than by rising $\Delta$, as the inset in Fig. \ref{fig:TeffnT} shows a smaller dispersion than the inset in Fig. \ref{fig:Teffn}. This can be explained by the initial thermal reservoir, which favors an incoherent evolution of the system. Nevertheless, the persistent frequency dependence hints a non thermal state.  We stress that the system does not reach a Gibbsian unique temperature state even after long times, as if it did, all the calculated effective temperatures should be equal and constant.



\section{Summary}
\label{sec:conclusions}

To conclude, we analyzed various dynamic correlation functions, for local and non-local operators after a quantum quench in an exactly solvable model in which the statistical description in terms of the GGE essentially leads to the emergence of thermal correlations. This is due to the existence of bi-partite eigenmode entanglement and a gap in the spectrum of the Hamiltonian that describes the initial state. For these correlations, the imposition of the FDT in the non-equilibrium context leads to the appearance of an effective temperature depending on frequency (and eventually momentum or position) that is different for each operator considered. Nevertheless, in the limit of strong initial entanglement, in agreement with the emergence of thermal behavior from the GGE, the local operators effective temperatures approach a well defined value (in a certain frequency region). However, the remaining frequency dependence of these temperatures and the fact that the non-local temperature does not follow this limit, discards thermalization to a standard Gibbs state in a strict sense. Finally, it is of particular interest the case of the frequency-dependent effective temperature obtained from the application of the FDT to the quasiparticle correlation function, evaluated at the dispersion relation of the Hamiltonian that performs the evolution. This effective temperature is directly related to the GGE Lagrange multipliers.

\acknowledgments

This work was partially supported by CONICET (PIP 0662), ANPCyT (PICT 2010-1907) and UNLP (PID X497), Argentina.

\bibliography{phys}

\end{document}